\documentclass[final,3p,times,twocolumn]{elsarticle}

\usepackage{amssymb}
\usepackage{multicol}
\usepackage{dcolumn}
\usepackage{graphicx}
\usepackage{mathtools}

\graphicspath{
	{./img/}
}

\journal{Acta Atsronautica}

\begin{document}

\begin{frontmatter}



\title{Combining Magnetic and Electric Sails for Interstellar Deceleration}


\author[mymainaddress]{{Nikolaos Perakis}
\corref{mycorrespondingauthor}}
\cortext[mycorrespondingauthor]{Corresponding author}
\ead{nikolaos.perakis@tum.de}

\author[mysecondaryaddress]{Andreas M. Hein}
\ead{andreas.hein@i4is.org}

\address[mymainaddress]{Technical University of Munich, Boltzmannstr. 15, DE85748 Garching, Germany}
\address[mysecondaryaddress]{Initiative for Interstellar Studies, 27-29 South Lambeth Road, London SW8 1SZ}

\begin{abstract}
The main benefit of an interstellar mission is to carry out in-situ measurements within a target star system. To allow for extended in-situ measurements, the spacecraft needs to be decelerated. One of the currently most promising technologies for deceleration is the magnetic sail which uses the deflection of interstellar matter via a magnetic field to decelerate the spacecraft. However, while the magnetic sail is very efficient at high velocities, its performance decreases with lower speeds. This leads to deceleration durations of several decades depending on the spacecraft mass. Within the context of Project Dragonfly, initiated by the Initiative of Interstellar Studies (i4is), this paper proposes a novel concept for decelerating a spacecraft on an interstellar mission by combining a magnetic sail with an electric sail. Combining the sails compensates for each technology’s shortcomings: A magnetic sail is more effective at higher velocities than the electric sail and vice versa. It is demonstrated that using both sails sequentially outperforms using only the magnetic or electric sail for various mission scenarios and velocity ranges, at a constant total spacecraft mass. For example, for decelerating from 5\% c, to interplanetary velocities, a spacecraft with both sails needs about 29 years, whereas the electric sail alone would take 35 years and the magnetic sail about 40 years with a total spacecraft mass of 8250 kg. Furthermore, it is assessed how the combined deceleration system affects the optimal overall mission architecture for different spacecraft masses and cruising speeds. Future work would investigate how operating both systems in parallel instead of sequentially would affect its performance. Moreover, uncertainties in the density of interstellar matter and sail properties need to be explored.  
\end{abstract}

\begin{keyword}
Magnetic Sail \sep Electric Sail \sep Interstellar Mission \sep Mission Design \sep Optimization



\end{keyword}

\end{frontmatter}


\section{Introduction}
\label{cha:intro}

The concept of manned and unmanned interstellar missions has been examined in different contexts by many authors within the past decades \cite{martin1979}. The main obstacle connected to the design of such a mission, is the necessity for an advanced propulsion system which is able to accelerate the spacecraft towards the target system within a reasonable time span. To overcome the vast interstellar distances, propulsion systems with high specific impulses, like the fusion based engines in the ICARUS and Daedalus projects have been proposed \cite{Long2009}, \cite{Daedalus}. Other methods rely on propellant-less systems like laser-powered light sails, as described in \cite{Forward1984}.

Accelerating a probe to high speeds and reaching the target system within short duration using advanced propulsion systems would be a big achievement for mankind. However, the scientific gain of an interstellar mission would be immensely increased with an extensive scientific payload. In order to produce valuable scientific results, the deceleration of the probe is required since it enables the study of star and planetary systems in detail \cite{Crawford2010}. For a more detailed analysis of exoplanets, involving surface operations, a deceleration down to orbital speeds is necessary.

Therefore, apart from the acceleration propulsion system, a further crucial mission component which is often overlooked, is the deceleration system of an interstellar mission. This has to demonstrate equally effective $\Delta v$ capabilities as the propulsion system. For that reason, methods utilizing propellant are not preferred, since they would induce large amounts of mass, which are inert during the acceleration and cruising phases of the mission. 

Two attractive concepts rely on utilizing magnetic and electric fields in order to deflect incoming ions of the interstellar space and thereby decelerate effectively. These systems called \textit{Magnetic Sail (Msail)} and \textit{Electric Sail (Esail)} were first proposed by Zubrin \cite{Zubrin1991} and Janhunen \cite{Janhunen2004} respectively. Since each one of those systems has a different design point and velocity application regime in which it performs optimally, the combination of the two can induce great flexibility in the mission design as well as better performance. To demonstrate these points, the example of a mission to Alpha Centauri is analyzed. This star system was chosen because it is the closest one to the earth at a distance of 4.35 light years and because the deceleration concept described in this paper, was inspired by the Dragonfly Competition of the i4is, which involved a light-powered light sail mission to Alpha Centauri \cite{Hein2016}. 

\section{Sail Properties}
\label{cha:sail_prop}

Before the comparison of the different deceleration methods takes place, the properties of each sail will be shortly analyzed and the assumptions used in the simulation of their performance will be explained. 

\subsection{Magnetic Sail (Msail)}
\label{ssec:msail}
The Msail consists of a superconducting coil and support tethers which connect it to the spacecraft and transfer the forces onto the main structure. The current through the coil produces a magnetic field. When the spacecraft has a non-zero velocity, the stationary ions of the interstellar medium are moving towards the sail in its own reference frame. The interaction of ions with the magnetosphere of the coil leads to a momentum exchange and a force on the sail, along the direction of the incoming charged particles. 

The force on the sail is calculated according to Equation \ref{eq:msail_force} \cite{Freeland2012}.

\begin{center}
\begin{equation}
F_{Msail}=0.345 \pi \left(m_p n_o \mu^{0.5} I R^2 v^2 \right)^{\frac{3}{2}}
\label{eq:msail_force}
\end{equation}
\end{center}

where $m_p$ is the mass of the proton, $n_o$ the number density of interstellar ions, $\mu$ the free space permeability, $I$ the current through the sail, $R$ its radius and $v$ its speed. Values for $n_o$ are proposed in \citep{Crawford2011} in the case of a space probe traveling to Alpha Centauri. In this work, a rather conservative value was implemented, with $n_o=0.03 \, \mathrm{cm^{-3}}$ corresponding to the expected values in the Local Bubble \citep{Crawford2011}. 

The main structural component introducing extra mass into the system is the sail itself, as well as its shielding and its deployment mechanism. The mass of the sail is defined by the maximal current density that can be achieved with the superconducting material, since this dictates the minimal cross sectional area for a specific current. According to Zubrin \cite{Zubrin1991}, the current densities of superconductors can reach up to $j_{max}=2 \cdot 10^{10} \mathrm{A/m^2}$ and this is the value used in the analysis.  For the material of the sail, the density of common superconductors like copper oxide (CuO) and YBCO was used, with $\rho_{Msail}=6000 \, \mathrm{kg/m^3}$.

The shielding mass required to protect the sail was modeled according to \citep{Daedalus}. This mass vaporizes due to collisions with interstellar atoms and ions and the total mass vaporized after time $T$ is given by Equations \ref{eq:shield_mass} and \ref{eq:shield_mass2}:

\begin{center}
\begin{equation}
m_{shield}=\int_0^T{\frac{\mathrm{d}m_{shield}}{\mathrm{d}t} \mathrm{d}t}
\label{eq:shield_mass}
\end{equation}
\end{center}

\begin{center}
\begin{equation}
\frac{\mathrm{d}m_{shield}}{\mathrm{d}t} =\frac{A_{ion}m_p n_o}{\Delta H} \frac{\beta c^3}{\sqrt{1-\beta^2}}\left[\frac{1}{\sqrt{1-\beta^2}}-1\right] 
\label{eq:shield_mass2}
\end{equation}
\end{center}

In Equation \ref{eq:shield_mass2}, $A_{ion}$ represents the cross sectional area of the coil, as seen from the direction of the incoming ions, $\Delta H$ is the vaporization enthalpy of the shielding material and $\beta=v/c$. Graphite was chosen as a shielding material since it combines a low density with high vaporization enthalpy. The shielding mass is calculated separately for each configuration, since its calculation  requires knowledge of the time-dependent profile for $\beta$. For that reason, its calculation is carried out with an iterative procedure. 

For the tether and support structures, a mass equal to 15 \% of the sail mass was used.

It is evident from the formula in Equation \ref{eq:msail_force}, that the magnetic sail is efficient for higher current values and larger dimensions. In the analyses presented in this work, the radius of the Msail was limited to 50 km. Although even larger dimensions can demonstrate better performance, it was thought that the deployment of bigger radii is far from the current or near-future technological capabilities and was therefore excluded from the analyses.

The main disadvantage of the magnetic sail is also evident when taking the force formula into account. At lower speeds, the force keeps getting reduced asymptotically, and hence the effect of the Msail at these velocities becomes negligible. This has as consequence that reaching orbital speeds (10-100 km/s) requires long deceleration duration. A magnetic sail would therefore be optimal for missions where no orbital insertion or surface operations in planetary systems are required but where a deceleration for prolonged measurements in the target system is sufficient. Its inefficiency in lower speeds indicates the need for a secondary system able to bring the velocity down to orbital values. 
 
\subsection{Electric Sail (Esail)}
\label{ssec:esail}
Similar to the Msail, where a magnetic field deflects incoming ions, the Esail uses an electric field to change the trajectories of the interstellar protons. The sail consists of extended tethers which are charged with a high positive voltage. 

The force on the Esail demonstrates a more complex dependency on the velocity compared to the Msail. The force can be described by Equation \ref{eq:esail_force} \cite{Janhunen2007}.

\begin{center}
\begin{equation}
F_{Esail}=N L \frac{3.09 \cdot m_p n_o v^2 r_o}
{\sqrt   
	{
	\exp {
	\left(
	\frac{m_p v^2}{e V_o}
	\ln{\left( 
	\frac{r_o}{r_w} 
	\right)} 
	\right)} -1}}
	\label{eq:esail_force}
\end{equation}
\end{center}

with $N$ standing for the number of tethers, $L$ their length, $V_o$ the voltage of the sail, $e$ the charge of the electron, $r_w$ the wire radius and $r_o$ the double Debye length $\lambda_{D}$, given by Equation \ref{eq:deb}:

\begin{center}
\begin{equation}
r_o=2\lambda_{D}=2\sqrt{\frac{\epsilon_o k_b T_e}{n_o e^2}}
\label{eq:deb}
\end{equation}
\end{center}

In the Debye length definition, $\epsilon_o$ is the electric permittivity of vacuum, $k_b$ the Boltzmann constant and $T_e$ the electron temperature of the interstellar plasma. $T_e$ was estimated according to \cite{Crawford2011}, so for the present analysis the value $T_e= 8000 \, \mathrm{K}$ was used. The wires were designed according to \cite{Janhunen2007}, with radius $r_w=5 \, \mathrm{\mu m}$ and density 1500 $\mathrm{kg/m^3}$

\begin{figure}[h]	
	\includegraphics[width=\linewidth]{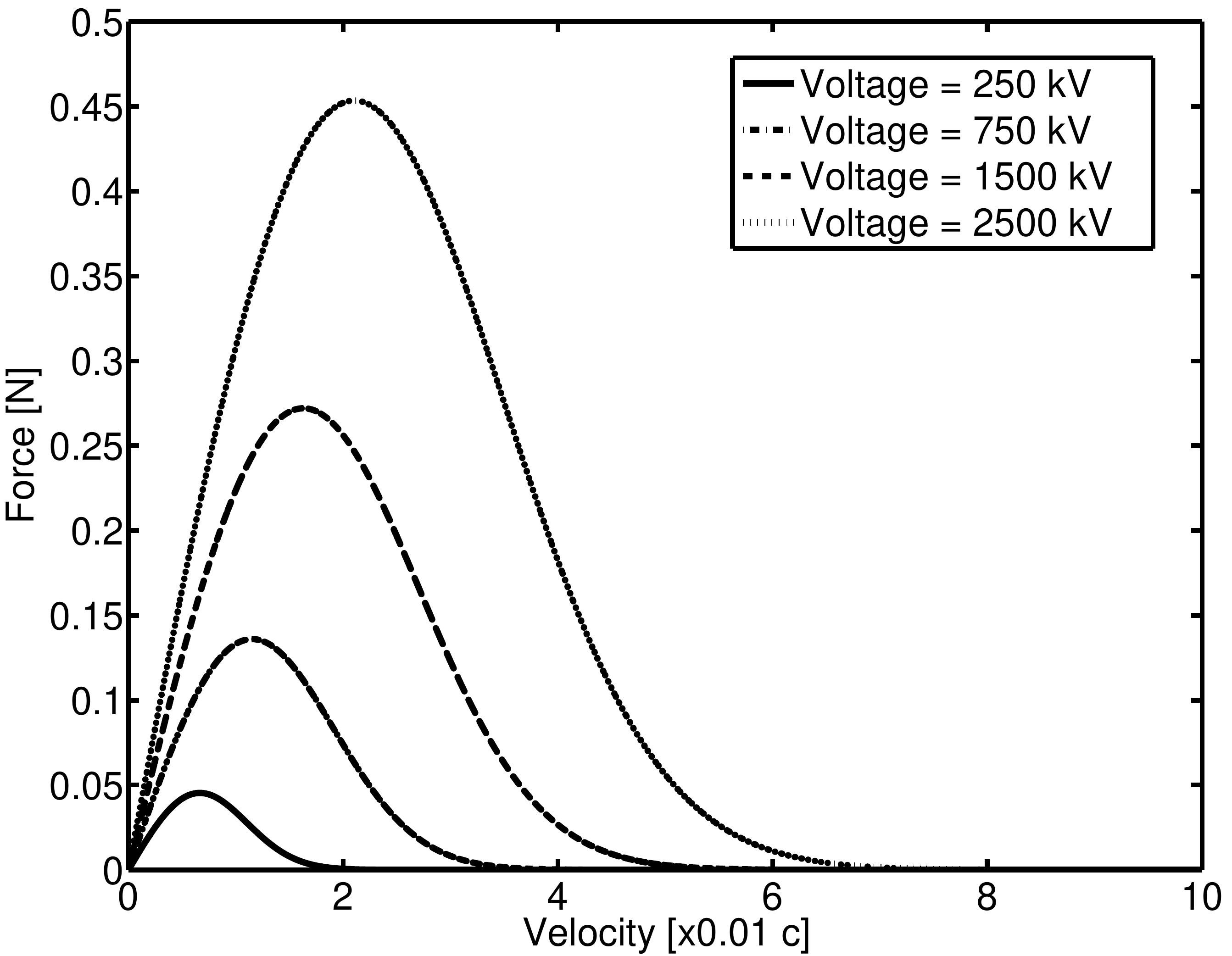}
	\caption{Force on an electric sail as a function of velocity}
	\label{fig:esail_force}
\end{figure}

It is evident from Equation \ref{eq:esail_force}, that the force increases proportionally to the number and length of the tethers as well as for a higher voltage. The dependency of the Esail force on the velocity of the probe however, displays a more complex character than the one for the Msail. Figure \ref{fig:esail_force} demonstrates this effect qualitatively for a constant total length of the tethers. It follows that the Esail is effective only within a region close to its maximal force. In order to decelerate a probe efficiently from high cruising speeds ($\geq 0.04 \,\mathrm{c} $) down to orbital values, a very high voltage is required according to Figure \ref{fig:esail_force}, or an increased total length of the tethers.

However an increase in tether length and voltage does not only imply a higher mass of the wires, but also a bigger power supply system. The positively charged tethers collide with the interstellar plasma electrons, which leads to a decrease of the voltage. In order to maintain the positive voltage of the wires, an electron gun has to be placed on board, leading to an additional mass for the power supply subsystem. The required power is described by Equation \ref{eq:power} \cite{Janhunen2007}:

\begin{center}
\begin{equation}
P= V_o \cdot I = 2 r_w V_o N L e n_o \sqrt{\frac{2e V_o}{m_e}}
\label{eq:power}
\end{equation}
\end{center}

with $m_e$ being the mass of the electron. The total mass of the Esail is put together from the mass of the tethers and the power system required for the operation of the electron gun. In the present work, the power system for the Esail was modeled with a specific power supply of 50 W/kg. Although the details of the power system were not part of this analysis, photovoltaic cells could be used, utilizing the laser beam power in combination with radioisotope thermoelectric generators and batteries. Another option is the use of electromagnetic tethers as an energy source, by means of electromagnetic induction as described in \cite{Matloff2005}.

It becomes clear that the Esail has a disadvantage when dealing with high speeds, because of the very high voltage and consequently system mass needed. For that reason, an additional system would be necessary for the initial deceleration from the high cruising speeds until the point where an optimally designed Esail can take over. 

\section{Combination of Msail and Esail}
\label{cha:combination}
After establishing the properties and the disadvantages of the individual sails in Section \ref{cha:sail_prop}, the benefits of combining the two subsystems for an effective deceleration in interstellar missions become clear.

Missions to neighboring star systems require high cruising speeds in order to reduce the total trip duration. There have been proposals based on fusion propulsion that aim to keep the total mission duration underneath 100 years \cite{Daedalus}, \cite{Hein2011}, which means that an average speed bigger than 0.0435 c is necessary in the case of Alpha Centauri \cite{Long2011}. The present analysis focuses on missions with the objective of performing scientific measurements in the target system, hence requiring orbital insertion around a star or a planet. In this context, the combination of Msail and Esail seems to be an elegant solution.  

Starting the deceleration phase of the mission with the use of a magnetic sail is beneficial as mentioned in Section \ref{ssec:msail}, due to the high forces produced in the large velocity range. As the velocity decreases, the force drops also and the Msail starts being ineffective. At this moment (which has to be optimally chosen as described later), the Msail can be switched off and detached from the spacecraft and the Esail can start operating. The electric sail must be designed to perform optimally in this velocity region and can decrease the velocity of the spacecraft further, until the required value for orbital insertion is achieved. The high flexibility of the tandem system comes in the expense of additional optimization effort. The two subsystems are dependent on each other and have to be designed simultaneously and an extra optimization parameter influences their design, namely the velocity value at which the start of operation for the Esail takes place.  

This idea resembles the concept of staging in conventional launchers with chemical engines. As soon as the first stage is done burning, it is detached, and the second stage, which has been optimally designed to operate in the higher altitude, is ignited. Similarly, as soon as the Msail reaches its weak performance point, it is dropped off and the Esail, which has been optimally designed to decelerate the remnant mass, starts operation.

The switching method presented in this paper is only one of the alternatives that can be realized with a combination of Msail and Esail. A further option would be that the Esail starts operation simultaneously with the Msail even at higher speeds, where it is not so effective. One would expect that this extra bit of braking force could improve the overall performance. This idea was not implemented in the present analysis, because the Esail tethers can be used for energy production according to \citep{Matloff2005} for the velocities that are far from their optimal design point. This way, instead of spending electric power for the operation of the Esail, which only has a small effect on the overall deceleration, the Esail can serve as a significant power supply source. 

Additionally, allowing the Msail to operate even at the velocity regime where it has lost its efficiency in parallel to the Esail instead of detaching it, would increase the decelerating force. However, the mass being decelerated would also increase and hence the magnitude of acceleration would not necessarily improve. A complete optimization model could include the start of operation of the Esail and the detachment of the Msail as two separate events. This brings some additional complexity to the model since it requires the optimization of a further parameter. However, it was examined for a single test case which is not in the scope of this paper and the obtained results showed a $<5 \%$ performance improvement, so it was ignored in this analysis. 

An extra benefit of ceasing the use of the Msail when the Esail starts operating, lies in utilizing the magnetically stored energy of the superconductor for the operation of the Esai. The current through the Msail could be discharged into batteries used for the power system of the electron gun before detachment, thereby turning the Msail to a Superconducting Magnetic Energy Storage \cite{Tixador2008}.

These considerations explain why a tandem switching method was preferred to a method where both systems run in parallel. It is easy to understand that the switching point should occur at a speed value where the acceleration with the Msail is equal to the acceleration with the Esail. Switching at a lower speed would imply that there is a time span where the probe is decelerating with a force smaller than what it could achieve by switching to the Esail and would become less effective. The same issue occurs for switching at higher speeds, since it means that the magnetic sail did not reduce the kinetic energy  by the amount it was optimally designed to.

\begin{figure}[h]
	\includegraphics[width=\linewidth]{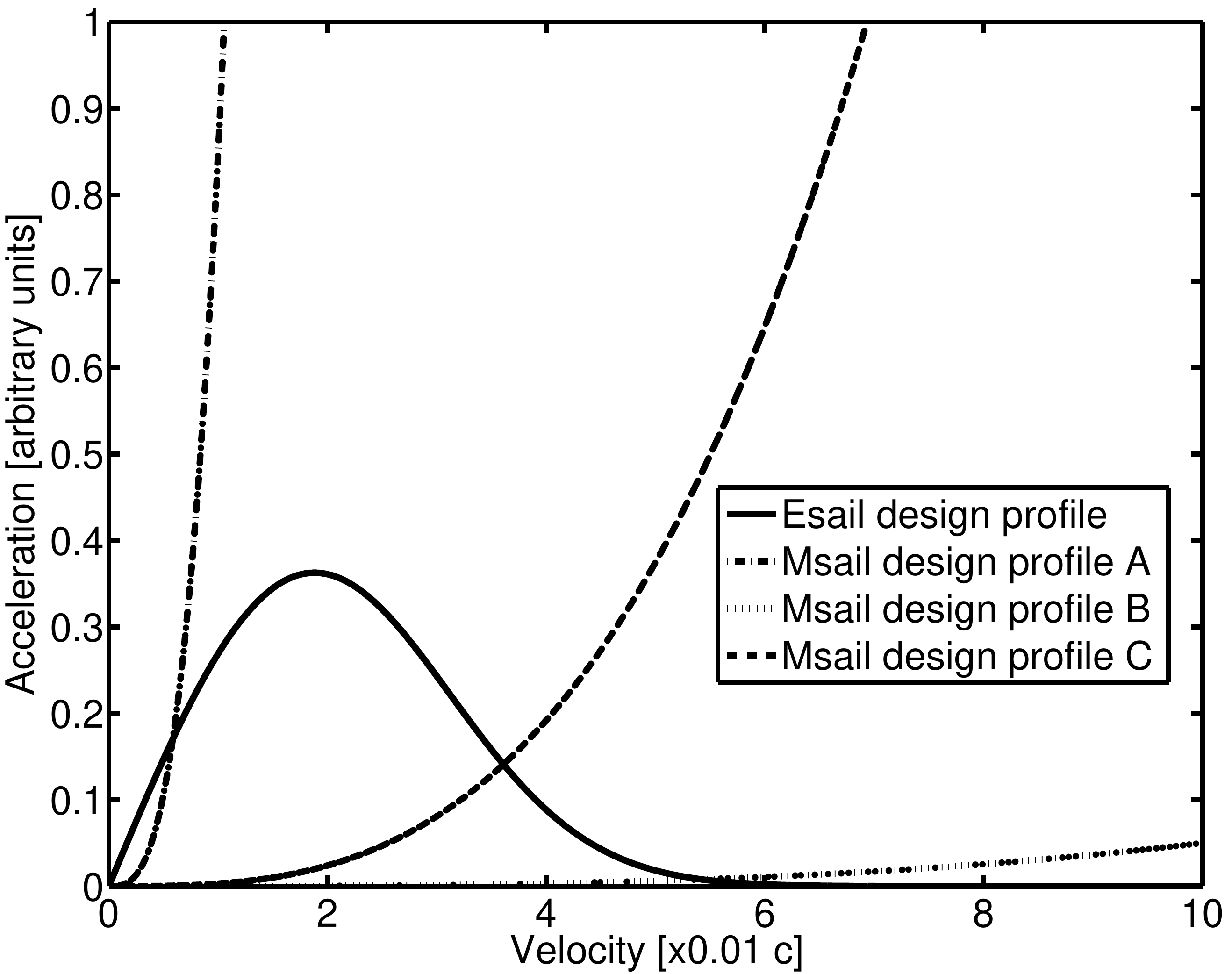}
	\caption{Qualitative description of Msail and Esail acceleration profiles}
	\label{fig:esail_msail_switch}
\end{figure}

The consideration of the optimal switching point between Esail and Msail can be qualitatively seen in Figure \ref{fig:esail_msail_switch}. 
In this image, a fixed design point for the Esail is chosen and an optimal design for the Msail is searched for. It is obvious, that the choice of an overly dimensioned Msail, like in the case of the design point A, is not very efficient. The intersection point of the acceleration profiles for Esail and Msail lies at velocities smaller than the point of the maximum Esail deceleration. Therefore, after the switch, the magnitude of acceleration would keep dropping and the highest Esail force would never be utilized. Although the acceleration magnitude would be bigger than what the Msail could have produced, the full potential of the Esail would still remain unused. 

In the case of profile B, the Msail is under dimensioned, hence leading to a high velocity for the switching point. At this regime, the Esail demonstrates a very low force and therefore does not reduce the speed of the probe efficiently. A significant time period has to elapse until the velocity reaches the optimal design point of the Esail, where the acceleration value is big enough to produce an effective braking of the spacecraft. 

Finally, case C seems to produce a better deceleration profile. The switching point lies in speeds higher than the optimal design point of the Esail. The Esail acceleration starts increasing immediately after the detachment of the Msail and is close to the optimal value, therefore utilizing the full potential of the electric sail, before starting to drop again.  

The combination of the two sails requires the optimization of the individual parameters for Msail and Esail (radius and current of superconductive loop, voltage, number and length of tethers) as well as of the velocity at which the operation of the Msail ceases.  

\section{Optimization process}
\label{cha:opt}
The optimization problem that was solved to come up with the optimal design of the deceleration system can be expressed as the minimization of the total deceleration duration $T_{decel}$. 

To determine $T_{decel}$ for a given sail configuration, the mass of the system and the force profile over time are necessary. 
The parameters $N$,$L$,$V_o$,$R$ and $I$ allow for the determination of $m_{Esail}$ and $m_{Msail}$. The combination of Msail and Esail requires the additional parameter of the switching velocity $v_{switch}$ which results to the force profile. Combining the mass and the force leads to the acceleration capabilities of the system. This way, the optimization parameters of the mathematical problem are summarized in Table \ref{tab:tab1} for the three deceleration methods.

For a given acceleration dependency on the velocity, $a(v)$, the total duration of the deceleration period (the cost function) is given by the expression in Equation \ref{eq:duration}:

\begin{center}
\begin{equation}
a(v)=\frac{\mathrm{d}v}{\mathrm{d}t} \Rightarrow \mathrm{d}t=\frac{\mathrm{d}v}{a(v)} \Rightarrow T_{decel}=\int_{v_{cruise}}^{v_{target}}  \frac{\mathrm{d}v}{a(v)}
\label{eq:duration}
\end{equation}
\end{center}

In the case of tandem deceleration, this takes the form of Equation \ref{eq:duration2}:

\begin{center}
\begin{equation}
\begin{multlined}
 T_{decel}=\int_{v_{cruise}}^{v_{switch}}  \frac{\mathrm{d}v}{a_{Msail}(v)} + \int_{v_{switch}}^{v_{target}}  \frac{\mathrm{d}v}{a_{Esail}(v)}\\ 
 =\int_{v_{cruise}}^{v_{switch}}  \frac{(m_{Msail}+m_{Esail}+m_{s/c})\mathrm{d}v}{F_{Msail}(v)} +\\ \int_{v_{switch}}^{v_{target}}  \frac{(m_{Esail}+m_{s/c})\mathrm{d}v}{F_{Esail}(v)}
 \end{multlined}
\label{eq:duration2}
\end{equation}
\end{center}
 and the objective of the minimization problem is summarized in:
 
 \begin{center}
 \begin{equation}
 T_{decel}=min!
 \end{equation}
 \end{center}

\begin{table}[h]
\begin{center}
  \begin{tabular}{ | c | c | }
    \hline
    Pure Msail & $I$, $R$ \\ \hline 
    Pure Esail & $N\cdot L$, $V_o$ \\ \hline
    Tandem Msail and Esail & $I$, $R$, $N\cdot L$, $V_o$, $v_{switch}$ \\ \hline  
    
  \end{tabular}
  \caption{Optimization parameters for each deceleration method}
  \label{tab:tab1}
  \end{center}
  \end{table}

Since the acceleration part of the mission is not captured in this analysis, the absence of any further constraints would shift the optimal solution to very high deceleration system masses. Since the performance of the system increases with increasing mass, an overly dimensioned Msail and Esail with infinite mass would minimize the cost function $T_{decel}$. When combined with the acceleration system however, such a large system would be inefficient since it would pose a large inert mass during the acceleration phase. For that reason, an additional constraint was introduced, namely an upper bound for the maximal deceleration mass. Hence this extra constraint was introduced as in Equation \ref{eq:constraint_m}:

\begin{center}
\begin{equation}
 m_{decel} \leq C
\label{eq:constraint_m}
\end{equation}
\end{center}

with $C$ being a predefined upper mass limit and 

\[
    m_{decel}= 
\begin{dcases}
    m_{Msail},& \text{for Msail deceleration } \\   
     m_{Esail},& \text{for Esail deceleration } \\
    m_{Msail} + m_{Esail},& \text{for tandem deceleration }
\end{dcases}
\]

Further constraints involve the initial and end velocity of the probe. This reads as in Equation \ref{eq:constraint_v}:

\begin{center}
\begin{equation}
 v(t=0) = v_{cruise} \; \text{and} \; v(t=T_{decel}) = v_{target}
\label{eq:constraint_v}
\end{equation}
\end{center}

This constraint is directly applied in the definition of the cost function $T_{decel}$, since it sets the limits of the integral calculation. 

In the case of the Msail and Esail combination, the switching velocity is to be modeled as well. One constraint for $v_{switch}$ is already present in Equation \ref{eq:duration2}, since it is set as the limit of the integral to be evaluated. Moreover, it has to be made sure, that the acceleration at the switching point between Msail and Esail remains continuous, as described in Section \ref{cha:combination}. Mathematically this yields:

 \begin{center}
\begin{equation}
\begin{multlined}
 a_{Msail}(v=v_{switch})=a_{Esail}(v=v_{switch}) \, \Rightarrow \\
\frac{F_{Msail}(v=v_{switch})}{m_{Msail}+m_{Esail}+m_{s/c}}=
 \frac{F_{Esail}(v=v_{switch})}{m_{Esail}+m_{s/c}}
\end{multlined}
\end{equation}
\end{center}

where $m_{s/c}$ stands for the spacecraft mass.

Moreover, as explained in Section \ref{cha:combination}, the switching point has to take place for velocities larger than the optimal operation point of the Esail and therefore:

 \begin{center}
\begin{equation}
v_{switch}>v(a_{Esail}=max)
\label{eq:constraint_switch2}
\end{equation}
\end{center}

Finally, the total deceleration distance $r_{decel}$ poses a further constraint. It has to be ensured, that there is sufficient distance available for the spacecraft to decelerate completely before it reaches Alpha Centauri. For that reason this should remain shorter than 4.35 light years. At the same time, there has to be some finite distance available for the acceleration and cruising phases, which are not part of the optimization and this was estimated equal to 1.5 light years. For that reason, the constraint was defined as in Equation \ref{eq:constraint_r}:

\begin{center}
\begin{equation}
 r_{decel} =
 \int_{v_{cruise}}^{v_{target}}  \frac{v\, \mathrm{d}v}{a(v)}
 \leq 2.85 \, \text{light years}
\label{eq:constraint_r}
\end{equation}
\end{center}

The cost function to be minimized ($T_{decel}$) is highly non-linearly dependent on the optimization parameters, and therefore linear programming methods were not useful. Moreover, due the lack of knowledge of the function gradient, the optimization took place with a pattern search method similar to the "direct search" proposed by Hooke \cite{Hooke}. This is the method utilized for all analyses in the present paper. 

After obtaining the optimal deceleration duration, the velocity and acceleration profiles as a function of time were calculated by means of numerical integration. A time propagation was implemented using a 4th order Runge-Kutta scheme, which served as a validation of the optimization results and provided a complete time profile of the spacecraft trajectory. 

\section{Results: Comparison of deceleration profiles}
\label{cha:resu}
Using the optimization method in Section \ref{cha:opt}, the performance of three separate deceleration methods was compared and the resuls are shown in this section. The three deceleration architectures are the following:

\begin{enumerate}
	\item Pure Msail deceleration
	\item Pure Esail deceleration
	\item Combination of Msail and Esail in tandem 
\end{enumerate}

In this test case, the mass of the spacecraft $m_{s/c}$ was chosen to be approximately equal to the launch mass of Voyager 1, so equal to 750 kg. Voyager is a space probe which was launched to perform flybys of Jupiter, Saturn and Titan and continued to explore the boundaries of the outer heliosphere \cite{Voyager1}. Since it is the only man-made probe so close to entering the interstellar space \cite{Voyager2}, it was considered relevant to calculate how its deceleration would look like in the case of a mission to another star system, requiring a deceleration phase. 

Only the deceleration phase of the mission was examined, so a cruising speed $v_{cruise}=0.05 \, c$ was chosen. The target speed was set to be equal to $v_{target}=35 \, \mathrm{km/s}$. This would correspond approximately to the orbital speed at a distance of 1 AU around Alpha Centauri A, which has a mass of 1.1 $M_{\odot}$ \cite{Centauri}. 

For each one of the three deceleration methods, an optimal design point was calculated in order to minimize the total deceleration duration $T_{decel}$. The mass of the deceleration system was restricted to be underneath 7500 kg, which corresponds to the tenfold spacecraft mass. A direct comparison is thereby possible, since all systems have the same effect on the acceleration phase and hence the overall mission design. 

At this point it has to be noted, that the restriction of the Msail radius described in Section \ref{ssec:msail} produces very week forces in the low speed limit (close to $v_{target}$), thereby resulting in duration close to 300 years. It was therefore dismissed from the calculations of pure Msail deceleration. The results shown here required a sail radius of 1000 km, which was considered to be unrealistic but was still included for completion. This demonstrates once again that the Msail as a standalone component is not sufficient for missions requiring orbital insertion in the target system.  

\begin{figure}[h]
	\includegraphics[width=\linewidth]{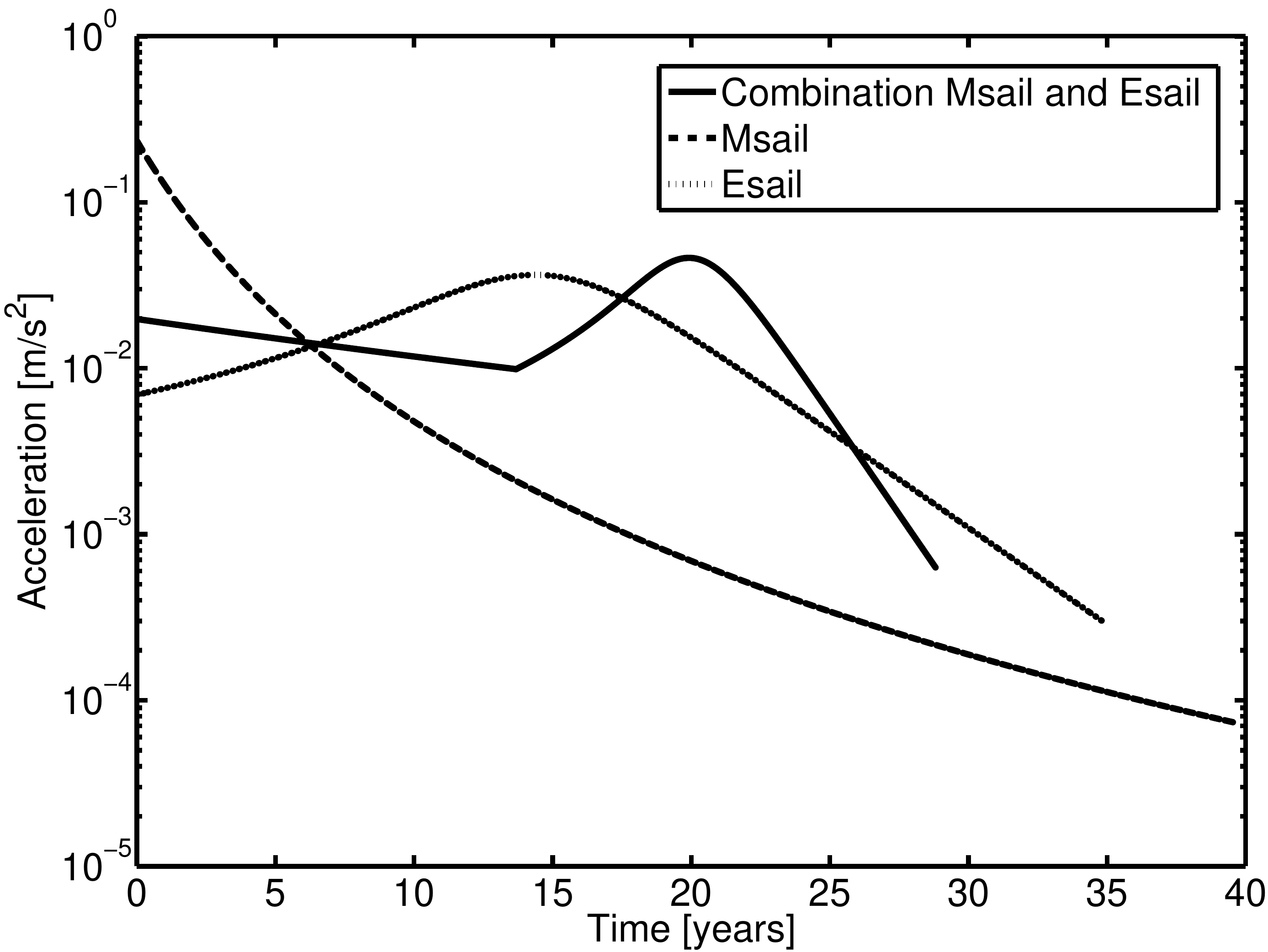}
	\caption{Comparison of deceleration methods: Acceleration profile over time}
	\label{fig:resu_a_vs_t_5}
\end{figure}

\begin{figure}[h]
	\includegraphics[width=\linewidth]{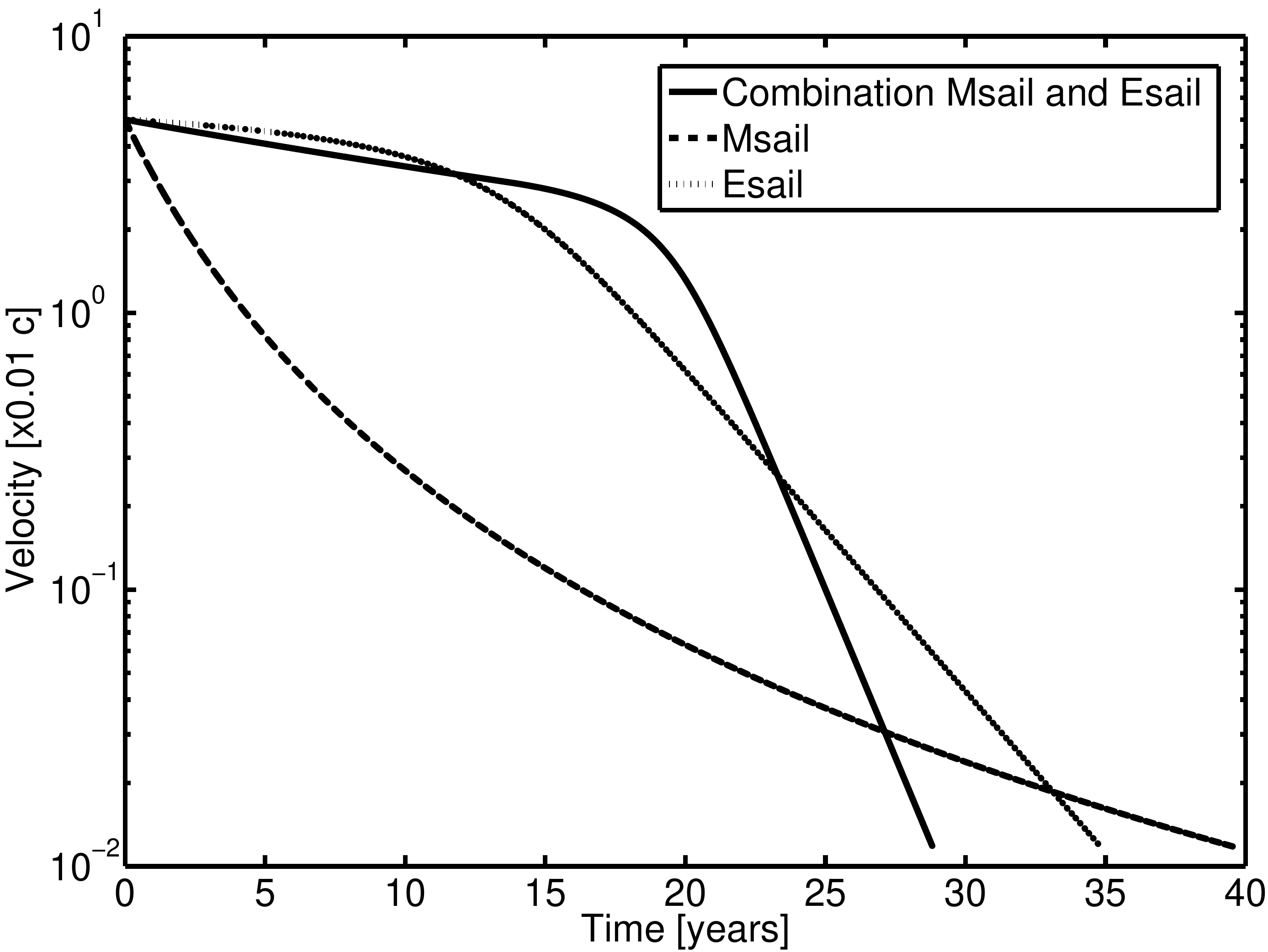}
	\caption{Comparison of deceleration methods: Velocity profile over time}
	\label{fig:resu_v_vs_t_5}
\end{figure}

The acceleration and velocity profiles over time are shown in Figure \ref{fig:resu_a_vs_t_5} and \ref{fig:resu_v_vs_t_5} respectively. Note that the curves in Figure \ref{fig:resu_a_vs_t_5} represent the magnitude of the acceleration, since the numeric values of acceleration are negative during the braking phase. The combination of the two sails requires 28.8 years as opposed to the 39.7 years of the Msail and the 34.9 years of the Esail. In the acceleration profile of the dual system, the discontinuity in the gradient represents the point where the switch between Msail and Esail takes place. This occurs after 13.67 years and at a speed  equal to approximately 0.03 c according to Figure \ref{fig:resu_v_vs_t_5}. This change is not detectable in the velocity profile, since the acceleration shows no discontinuity during the switch from the one system to the other, leading to a smooth velocity curve. 

Initially, the acceleration of the Msail method is the highest. This makes sense because the magnetic sail used in the tandem method is smaller than in the pure Msail method, in order to satisfy the equal mass requirement. After some time however, the magnitude of the acceleration in the tandem method becomes larger and eventually leads to a smaller duration. 

At this point, it is also important to mention that the pure Msail method is the optimal solution when a higher target speed is needed. Figure \ref{fig:resu_v_vs_t_5} demonstrates this effect since the velocity curve of the Msail is lower than the other two for the whole duration apart from the lower velocity range, where it flattens. The absence of orbital insertion (leading to $v_{target}$ being an order of magnitude larger), would therefore make the Msail the most effective solution. 

This test case demonstrates the potential that a combination of Msail and Esail has in the design of an interstellar mission, since it outperforms each individual system in particular mission configurations. However, during a complete mission design, the minimal deceleration duration is not the only parameter to be optimized and the interaction of the deceleration system with the other components (influence on acceleration, effect of deceleration distance) has to be taken into account. 

\section{Interaction with mission design}
After having established that the method of tandem deceleration with Msail and Esail can bring benefits to the total duration of the deceleration phase before orbital capture, it is interesting to determine how this system interacts with the acceleration and cruising phases. 

\subsection{Influence of cruising velocity}
\label{cha:compB}

In Section \ref{cha:combination}, a single value for the cruising speed was examined. In this section, the effect of a variable cruising speed on the design characteristics of the tandem deceleration system is presented. 

For this analysis, two different spacecraft masses are compared. Apart from the Voyager-like spacecraft introduced in Section \ref{cha:combination}, the profile of a heavier vehicle with $m_{s/c}=4000 \, \mathrm{kg}$ is calculated. This value was chosen since it is approximately equal to the launch mass of the Mars Science Laboratory (MSL). This robotic space probe was sent to Mars and included a rover with a landing system and instruments for biological, geochemical and geological measurements on the surface of the planet \cite{Mars}. Since a similar mission to an exoplanet would be of high scientific value \cite{Hein2011}, an MSL-like spacecraft was used. The restriction for the total mass of the deceleration system being maximally ten times the spacecraft mass was maintained. 

\begin{figure}[h]
	\includegraphics[width=\linewidth]{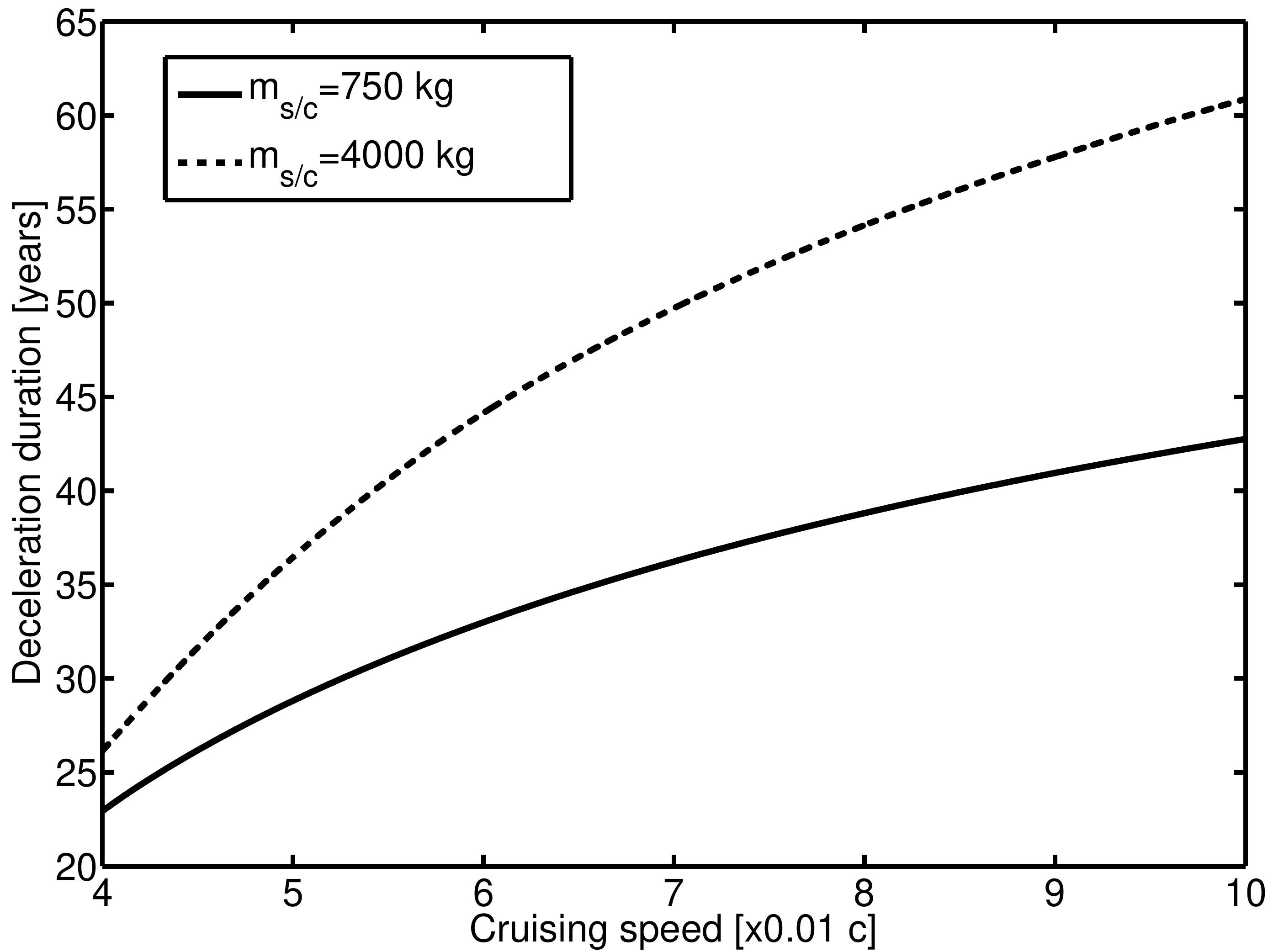}
	\caption{Deceleration duration of optimal configuration as a function of the cruising velocity}
	\label{fig:durationB}
\end{figure}

\begin{figure}[h]
	\includegraphics[width=\linewidth]{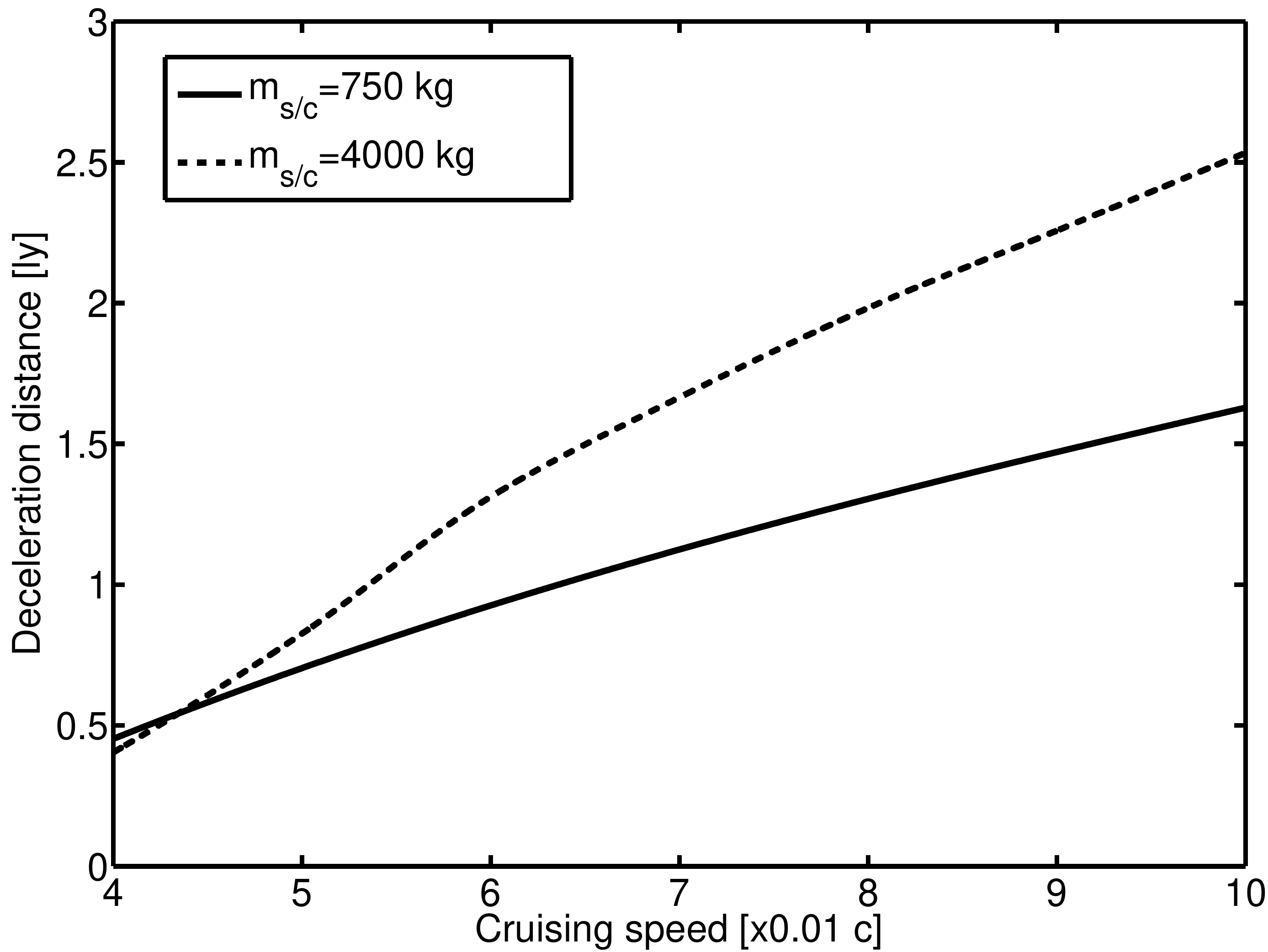}
	\caption{Deceleration distance of optimal configuration as a function of the cruising velocity}
	\label{fig:distanceB}
\end{figure}

Figures \ref{fig:durationB} and \ref{fig:distanceB} show the dependency of the deceleration duration and distance on the cruising speed. It is intuitive that a larger initial speed requires a larger deceleration duration, since the total $\Delta v$ that has to be provided by the deceleration system increases. 

The same occurs for the deceleration distance, as Figure \ref{fig:distanceB} demonstrates. A higher spacecraft mass also increases the inertia of the system during deceleration and hence the time and distance required. An important indirect result stemming from Figure \ref{fig:distanceB} is that high cruising speeds are not always optimal for a minimal mission duration. In the case of the 4000 kg spacecraft, a 0.1 c cruising speed leads to a deceleration distance close to 2.5 light years. When taking into account that the distance to Alpha Centauri is 4.35 light years, one deduces that there are only 1.85 light years available for the acceleration and cruising phases. However, the buildup of such a high speed could require a larger acceleration distance depending on the propulsion system. Therefore, reaching such a high speed in a mission to Alpha Centauri may not be necessary or useful, due to the extreme deceleration distance connected to it.

\begin{figure}[h]
	\includegraphics[width=\linewidth]{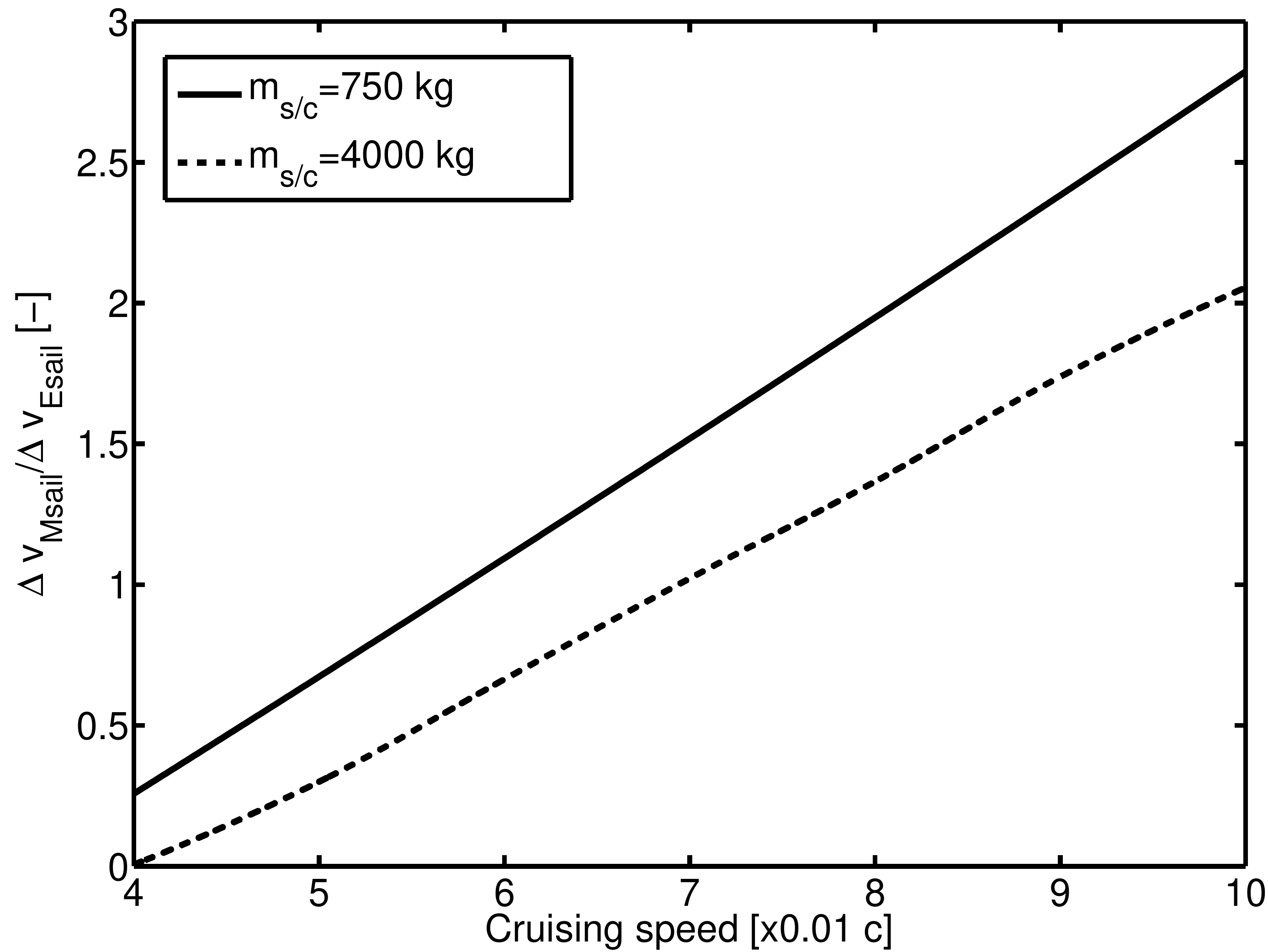}
	\caption{Optimal mass ratio of Msail to Esail as a function of the cruising velocity}
	\label{fig:vratioB}
\end{figure}

\begin{figure}[h]	
	\includegraphics[width=\linewidth]{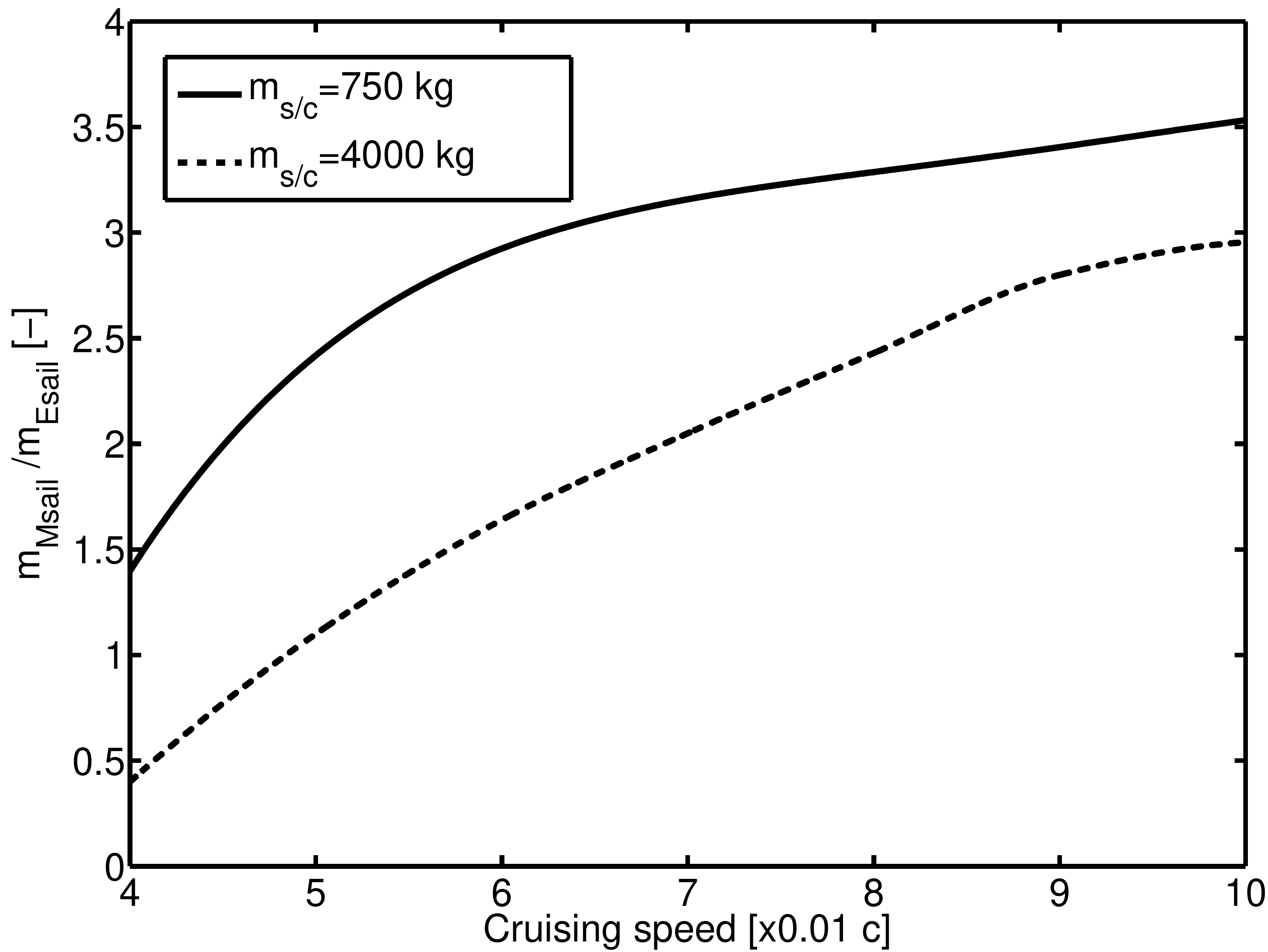}
	\caption{Optimal $\Delta v$ ratio of Msail to Esail as a function of the cruising velocity}
	\label{fig:mratioB}
\end{figure}

The mass and velocity change distribution between Msail and Esail are also interesting to examine as a function of the cruising speed. Figure \ref{fig:mratioB} shows the ratio of the Msail mass $m_{Msail}$ to the Esail mass $m_{Esail}$ and Figure \ref{fig:vratioB} the ratio of the velocity changes $\Delta v_{Msail}$ and $\Delta v_{Esail}$ at the optimal configuration for each crusing speed. 

The velocity change ratio demonstrates a nearly linear profile in Figure \ref{fig:vratioB}, which increases with the cruising speed. This can be explained with the good performance of the Msail in higher speeds. Since the Msail is efficient in the high speed regime, it is logical that it will also take over most of the deceleration. Moreover, the results show that a higher spacecraft mass leads to a lower $\Delta v$-ratio. 

Since the velocity changes are proportional to the mass of each subsystem, it is expected that the mass ratio also increases with the cruising speed, as shown in Figure \ref{fig:mratioB}. In this case however, the increase in mass ratio tends to be slower and resembles a logarithmic growth. 

The results show a general preference towards the Msail deceleration for higher cruising velocities which is reflected in the $\Delta v$ and mass distribution of the deceleration system. 

\subsection{Effect of deceleration system mass}
\label{cha:compC}

The deceleration system is an integral part of the mission design and cannot be analyzed independently of the acceleration phase when an interstellar mission is being developed. The main effect that the deceleration system has on the acceleration phase is its mass, which needs to be accelerated as well. Therefore, a deceleration system which is as light as possible but still produces the necessary $\Delta v$ change in short amount of time and in short distance is required. 

\begin{figure}[h]	
	\includegraphics[width=\linewidth]{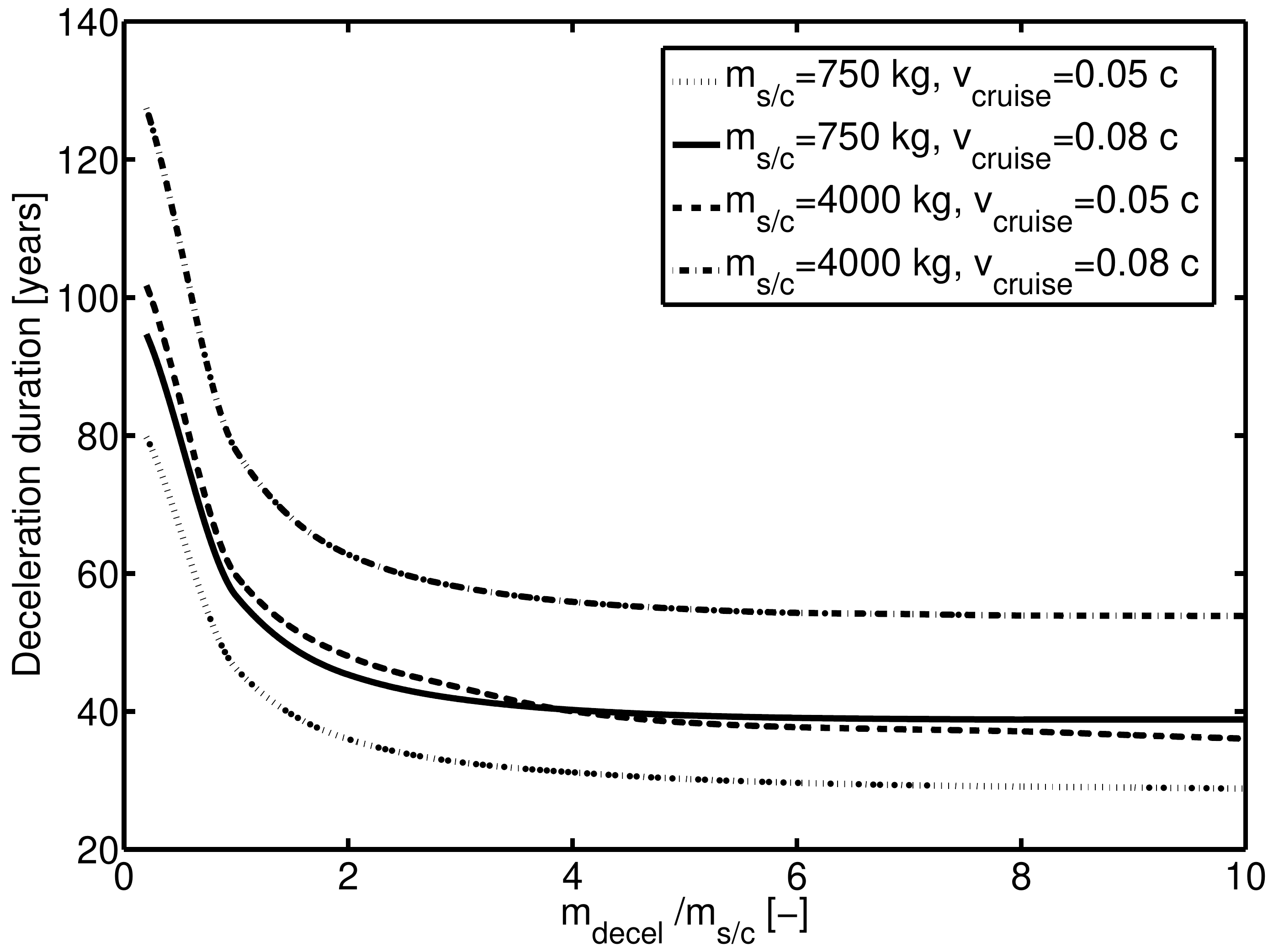}
	\caption{Optimal deceleration duration as a function of the deceleration system mass}
	\label{fig:durationC}
\end{figure}

The effect of the tandem deceleration system mass on its performance was examined. In the previous sections, the requirement of the deceleration system mass being smaller than ten times the spacecraft mass was utilized. This boundary condition was introduced so that an easier comparison between different configurations could take place. In the present analysis however, the ratio between deceleration system mass and spacecraft mass was varied. The two spacecraft masses described in Section \ref{cha:compB} as well as two different cases for the cruising speed were compared to each other. Figure \ref{fig:durationC} shows the results. 

An increased mass of the deceleration system leads, as expected, to a shorter deceleration duration. It is however notable, that the curves tend to saturate for larger masses. This implies that a larger deceleration mass, although having a great impact on the design of the acceleration phase because of additional inertia, only provides a small benefit to the overall deceleration performance. Quantitatively, taking the example of the 4000 kg spacecraft with 0.08 c cruising speed in Figure \ref{fig:durationC}, one observes that a mass ratio of 10 leads to a minimal duration equal to 53.83 years whereas a mass ratio equal to 4 results in 55.90 years. Hence an increase of 150 \% in the mass of the deceleration system, produces only a 3.7 \% increase in the performance of the system. This trend is maintained for all configurations and it is evident, that when the complete mission is designed and all mission phases are optimized simultaneously, deceleration system masses are preferred, which are further from the saturation limit and still produce sufficient performance.

\section{Conclusion}
\label{cha:conclusion}
Magnetic and electric sails have been proposed as propulsion systems for interstellar and interplanetary missions. In the case of interstellar missions with short trip duration and need for orbital insertion around a target system, each one of these sails demonstrates some disadvantages: Msails fail to produce sufficient forces in the low speed limit and Esails require very large masses in order to decelerate from the high cruising speeds of interstellar missions.

The present paper demonstrated that a combination of the two systems in tandem (initial deceleration with Msail and following braking with Esail) can have a better performance in certain configurations. Small unmanned missions were examined in this context and a generalization of this method for manned missions with larger spacecraft masses would be interesting since it would show the applicability limits of the system.
The combination of the two systems in series is not the only method that could improve the deceleration characteristics. Although this was the main architecture analyzed in the paper, operation of the two sails in parallel should also be further examined and controlled for additional increase in performance.

The overall design of an interstellar mission requires the optimization of the deceleration system not as a standalone component, but simultaneously with the main propulsion system of the acceleration phase and with the design of the cruising phase. The flexibility of the combination of the two sails includes further optimization parameters in the mission architecture, since the switching point between Msail and Esail deceleration has to be also optimized for maximal performance. 

Finally, the technical design of each sail, including the chosen density of the materials, power system, shield masses etc. as well as parameters with uncertainty, like the properties of the interstellar plasma, influence the optimal solution and should be carefully treated when an interstellar mission is being designed, because they directly affect the deceleration performance and consequently the overall mission architecture.

\section{Acknowledgments}
\label{cha:ack}
The authors would like to thank the Initiative for Interstellar Studies for organizing the Project Dragonfly Competition, which gave the inspiration for the present study. Moreover, the authors would like to thank the members of the WARR Interstellar Spaceflight Team at the Technical University of Munich, Johannes Gutsmield, Artur Koop, Martin J. Losekamm and Lukas Schrenk for their ideas during the design of the Dragonfly mission.





\section*{References}
\bibliographystyle{elsarticle-num} 
\bibliography{TandemBib}



\end{document}